\documentclass[conference]{IEEEtran}

\makeatletter
\def\ps@headings{%
\def\@oddhead{\mbox{}\scriptsize\rightmark \hfil \thepage}%
\def\@evenhead{\scriptsize\thepage \hfil \leftmark\mbox{}}%
\def\@oddfoot{}%
\def\@evenfoot{}}
\makeatother
\newcommand{\resetcounters}{\setcounter{equation}{0}}
\usepackage{amsfonts,amsmath,amstext,verbatim}
\usepackage{amssymb}%
\usepackage{setspace}
\usepackage{ifpdf}
\makeatother  \renewenvironment{abstract}{%
  \small\bfseries\textit{Abstract}:  }

\ifCLASSINFOpdf
\usepackage[pdftex]{graphicx}
\usepackage[pdftex,colorlinks,%naturalnames,%
           bookmarks=true,%
           pdftitle=...,%
           pdfauthor=B.\ Blaszczyszyn%
           \ -\  P.\ Muhlethaler,]{hyperref} 
\usepackage[numbers,sort&compress]{natbib}
\usepackage{hypernat}
\else

 \usepackage[dvips]{graphicx}
\usepackage[numbers,sort&compress]{natbib}
\fi

\newcommand{\LD}{\mathbf{L}}

\newcommand{\md}{\text{\rm d}}
\newcommand{\ir}{{\mathbb{R}}}
\newcommand{\E}{{\mathbf E}}
\newcommand{\Pro}{{\mathbf P}}
\newcommand{\ind}{1\hspace{-0.30em}{\mbox{I}}}

\newtheorem{Th}{Theorem}[section]
\newtheorem{prop}[Th]{Proposition}
\newenvironment{Prop}{\bf\begin{prop}\rm\em}{\end{prop}} % proposition
\newtheorem{cor}[Th]{Corollary}
\newenvironment{Cor}{\bf\begin{cor}\rm\em}{\end{cor}} % proposition
\newtheorem{res}[Th]{Result}
\newtheorem{lemma}[Th]{Lemma}
\newenvironment{Lemma}{\bf\begin{lemma}\rm\em}{\end{lemma}} % lemma
\newtheorem{fact}[Th]{Fact}
\newtheorem{exe}[Th]{Example}
\newenvironment{Exe}{\bf\begin{exe}\rm}{\end{exe}} % Fact  in italic
\newtheorem{remark}[Th]{Remark}
\newenvironment{Remark}{\bf\begin{remark}\rm}{\end{remark}} % Fact  in italic

\newcommand{\rem}{{\medskip\noindent \bf Remark:~\nolinebreak}}

\newcommand{\argmin}{\mathop{\text{\rm arg~min}}}

\newcommand{\bfF}{\mathbf{F}}

\newcommand{\calS}{{\mathcal{S}}}
\newcommand{\calD}{{\mathcal{D}}}
\newcommand{\calL}{{\mathcal{L}}}

\newcommand{\calT}{{\mathcal{T}}}

\makeatother  \renewenvironment{abstract}{%
  \small\bfseries\textit{Abstract}:  }

\begin{document}
\title{A New Phase Transition for Local Delays in MANETs}

\author{Fran{\c c}ois Baccelli ({\em  INRIA/ENS}) and
 Bart{\l}omiej B{\l}aszczyszyn ({\em INRIA/ENS and
   Math. Inst. Univ. of Wroc{\l}aw})
}

\maketitle
\thispagestyle{empty} \pagestyle{empty}

\begin{abstract}
We study a slotted version of the Aloha Medium Access (MAC) protocol
in a Mobile Ad-hoc Network (MANET). Our model features transmitters
randomly located in the Euclidean plane, according to a Poisson point
process and a set of receivers representing the next-hop from every
transmitter. We concentrate on the so-called outage scenario, where a
successful transmission requires a Signal-to-Interference-and-Noise
(SINR) larger than some threshold. We analyze the local
delays in such a network, namely the  number of times slots
required for nodes to transmit a packet to their prescribed next-hop receivers. 
The analysis depends very much on the receiver scenario and on the 
variability of the fading. In most cases, each node has 
finite-mean geometric random delay and thus a positive next hop throughput.
However, the spatial (or large population) averaging of these individual
finite mean-delays leads to infinite values in several practical cases,
including the Rayleigh fading and positive thermal noise case.
In some cases it exhibits an interesting phase transition phenomenon
where the spatial average is finite when certain model parameters 
(receiver distance, thermal noise, Aloha medium access probability)
are below a threshold and infinite above.
To the best of our knowledge, this phenomenon, which we propose
to call the wireless contention phase transition, has not been discussed
in the literature.
We comment on the relationships between the above
facts and the heavy tails found in the so-called ``RESTART''
algorithm. We argue that  the spatial average of the 
mean local delays is infinite primarily because of
the outage logic, where one transmits full packets at time slots when
the receiver is covered at the required SINR and where one wastes all
the other time slots. This results in the ``RESTART''
mechanism, which in turn explains why we have infinite
spatial average. Adaptive coding offers another nice way of breaking the
outage/RESTART logic. We show examples where the average delays are
finite in the adaptive coding case, whereas they are infinite in the
outage case.
\end{abstract}


\section{Introduction}
\resetcounters
Aloha is one of the most common examples of a multiple access 
protocol \cite{Schwartz87,Bertsekas01}.
The classical approach to Aloha adopts simplified packet
collision models in which simultaneous
transmissions are never successful. This makes this classical approach not well
adapted to a wireless MANET scenario, where it is the SINRs at
different receiver locations 
which determine the set of successful transmissions.
The present  paper contributes to the study of Spatial Aloha, a variant of Aloha well
adapted to MANETs. More precisely, it bears on the mathematical
analysis of Spatial Aloha in the context of large Mobile
Ad hoc Network (MANETs) with randomly located nodes. It focuses 
on the SINR coverage scheme, where each transmission requires
that the receiver be covered by the transmitter with a minimum SINR.
The present paper uses the stochastic geometry approach proposed in \cite{allerton,jsac} 
where the {\em space-time density of successful transmissions} was evaluated and optimized. 
The present paper identifies a potential weakness of this SINR coverage scheme: in most practical 
cases, and in particular as soon as thermal noise is bounded from
below by a positive constant, the mean delay for a typical
node to transmit a packet is infinite e.g. in the Rayleigh fading case.
%In MANETs the wireless nodes are 
%randomly displayed and stochastic geometry can be used
%to provide closed form expressions for various kinds of spatial
%averages pertaining to the performance analysis of
%this MAC protocol. 
\subsection{Main Paper Contributions}
The main {\em modeling advances} of the present paper are two-fold:
\begin{itemize}
\item  We add a {\em time dimension} to the existing spatial analysis. 
Time is slotted, and we hence focus on slotted Aloha. 
We assume that the geographical locations of the MANET
nodes remain unchanged over time and that only the variables modeling the
MAC status (allowed to transmit or delayed) and the channel characteristics
(such as fading and thermal noise) vary over time.
In other words, we consider a full separation of the time scale of node
mobility on one side and the time scale of MAC and physical layer on the
other side, which makes sense in many practical situations, where the
former is much larger than the later. This is in 
contrast with what happens in delay tolerant networks (DTNs) ; see
e.g.~\cite{GrossglauserTse2002,Jacquetinfo09}.
%Let us stress that this time dimension is {\em not} that of
%delay tolerant networks (DTN), where one leverages node mobility
%to contribute to the transport of packets (see
%e.g. \cite{GrossglauserTse2002} 
%and \cite{Jacquetinfo09}).

\item We propose some more {\em realistic receiver models} than the 
fixed-distance receiver model introduced in \cite{allerton,jsac};
the new models are inspired by the fact that the routing schemes typically choose, as next-hop, 
the closest possible receiver among some common set of potential receivers.
%  this   goes beyond the  assumption 
%that the receivers are within a fixed  distance from their emitters.
\end{itemize}
The main theoretical advances of the paper bear on the analysis of the {\em local
  delays} in such MANETs; the local delay of a node is the random numbers of
slots required by this node to successfully transmit a packet to its
next-hop node.   
We first perform a time analysis of these local delays given the
location of the MANET nodes. This analysis shows that each node has 
finite-mean geometric conditional random delay and thus a positive 
throughput to its next-hop receiver.

However, the spatial irregularities of the network imply that
these conditional throughputs vary from node to node; in a Poisson
configuration, one can find nodes which have an arbitrarily
small throughput and consequently an arbitrarily large delay.
In order to capture the performance of the whole MANET, one usually
considers  its ``typical node''. The typical node statistics 
are spatial (or large population) averages of the individual node
characteristics.
Our analysis shows in several practical cases,
including the Rayleigh fading and positive thermal noise case, 
that the local delay of the typical MANET node 
is heavy tailed and that its mean is infinite,
 which however does {\em not} imply that the mean throughput of the
 typical node is null.   Moreover, in certain cases,  
the mean local delay of the typical node exhibit a phase transition
phenomenon that we propose
to call the {\em wireless contention phase transition}: it is 
finite when certain model parameters (as receiver distance, thermal
noise, Aloha medium access probability) 
are below a threshold and infinite above.

On the theory side, we also comment on the connections between the
heavy tailedness alluded to the local delay of the typical MANET node
and the observations of
\cite{JelenkovicTanInfocom07,JelenkovicTan2007}, 
where it was shown that a finite population ALOHA model with
variable and unbounded size packets has power law transmission
delays. Although the 
physical phenomena at hand are quite different here (our spatial MANET
model has fixed-packet-sizes) and there, we use 
the so-called ``RESTART algorithm phenomenon'', which has recently received 
a lot of attention (see e.g.~\cite{Asmussen,JelenkovicTan2009}), to establish
some links between our findings and the results of \cite{JelenkovicTanInfocom07,JelenkovicTan2007}.
More precisely, we argue  that, in our spatial MANET model with fixed-packet-size Aloha MAC, 
the delay of the typical node is heavy tailed and can have infinite mean due to a ``RESTART'' phenomenon, 
where the {\em spatial irregularities in the MANET} play the same role as
the packet size variability in \cite{JelenkovicTanInfocom07} and \cite{JelenkovicTan2007}. 

The main practical contributions of the paper bear on ways to
guarantee finite mean local delay of the typical node
by {\em increasing diversity} in the MANET.
The proposed solutions have the potential of
breaking the RESTART rigidity, e.g. by increasing
the variability of fading, by increasing mobility, by
adding appropriate receivers or, finally, by using adaptive coding,
which completely brakes out  the
outage/RESTART logic.

The paper is organized as follows. In the remaining part of this
section we briefly present the related work. 
We present our MANET model in Section~\ref{s.MANET}. 
The mean local delays are introduced and evaluated in Section~\ref{s.TheLD}.
We analyze the phase-transition phenomenon in Section~\ref{s.Phase-Transition}. 
Section~\ref{s.FiniteMean} focuses on the ways to make the mean delays finite.
We conclude our work in Section~\ref{s.conclusion}.

\subsection{Related Work}
\label{ss.RelatedWork}
As already mentioned, the present paper assumes a full time-scale
separation for the mobility on one side and for the MAC and physical layer
on the other side. This assumption makes a major difference between what
is done in this paper and what is done in DTNs, where one leverages node mobility
to contribute to the transport of packets. There is a large number of publications
on the throughput in DTNs and we will not review the literature on the
topic which is huge.
Let us nevertheless stress that there are some interesting connections between the
line of though started in (\cite{GrossglauserTse2002}), where it was first shown that
mobility increases capacity and what is done in Section~\ref{secalo:HMN}.
We show in this section that mobility helps in a way which is quite different from
that considered in \cite{GrossglauserTse2002}:
mobility may in certain cases break dependence and hence mitigate the RESTART phenomenon,
it may hence decrease the mean local delay of the typical node (or equivalently increase
its throughput), even if one {\em does not use mobility to transport packets}. 

Among recent papers, that we are aware of, on the  time-space analysis of MANETs,
we would quote \cite{Jacquetinfo09} and \cite{Haenggi-spaswin09}. The former focuses
on node motion alone and assumes that nodes within transmission range can transmit
packets instantaneously. The authors then study the speed at which some multicast
information propagates on a Poisson MANET where nodes have independent motion
(of the random walk or random way-point type).
The latter focuses on a first passage percolation problem,
however, the model used in \cite{Haenggi-spaswin09} is the so called protocol model; 
its analysis significantly differs from that of our physical (SINR-based) model.
In particular, there is no notion of local delay.

\section{MANET model with Slotted Aloha}
\resetcounters
\label{s.MANET}
\subsection{Space-Time Scenarios}
\label{ss.ST-scenarios}
In what follows we describe a few space-time models considered in this
paper. In the most simple case this will consist in adding 
the  time-dimension to the Poisson Bipolar model introduced and
studied in~\cite{jsac}. We will go however beyond the simple receiver
model proposed there.
%of Section~\ref{ss.Bipolar} and to its
%extensions considered in Section~\ref{s.beyond.dipole}
%(such as the independent Poisson receiver (INR) model, 
%the MANET nearest receiver (MNR) and
%the nearest neighbor (MNN) model, as well as the multicast model.
The idea consists in assuming 
that the geographical locations of the  MANET nodes remain unchanged over time
and that the MAC status (allowed to transmit or delayed) 
and other characteristics of each node 
(as fading and thermal noise) vary over time.
Time is discrete with a sequence of {\em time slots}, $n=0,1,\ldots$,
w.r.t. which all the nodes are perfectly synchronized.

More precisely, we assume that a snapshot of the MANET 
can be represented by a marked Poisson point process
(P.p.p.) on the Euclidean plane, where the point process is homogeneous
with intensity $\lambda$ and represents the random locations of the nodes
and where the multidimensional mark of a
point/node carries information about its MAC status and other characteristics of the 
channels in the successive time slots.
This marked Poisson p.p.  is denoted by
$\widetilde\Phi=\{(X_i,(e_i(n),y_i,\bfF_i(n), W_i(n):n)\}$, where:
\begin{itemize}
\item $\Phi=\{X_i\}$ denotes the {\em locations of the potential
transmitters}  of the MANET; $\Phi$ is always assumed Poisson with positive and finite intensity $\lambda$;
\item\label{i.ei}
 $e_{i}(n)$ is the {\em MAC decision} of point $X_i$ of $\Phi$ at time $n$;
we will always assume that, given $\Phi$, the random variables
$e_i(n)$ are i.i.d. in $i$ and $n$; i.e. 
in space and time, with $\Pr\{\,e_i(n)=1\,\}=1-\Pr\{\,e_i(n)=0\,\}=p$.    
\item\label{i.yi} $y_i\in\ir^2$ is the location of the 
{\em receiver} of the node $i$; a few scenario for the choice of the
receivers are presented in Section~\ref{ss.receiver} below.
\item\label{i.Fi}
 $\bfF_i(n)=\{F_i^j(n):j\}$ is the  {\em virtual power} 
emitted 
by node $i$ (provided $e_i=1$) 
towards receiver $y_j$ at time $n$.
By virtual power $F_i^j(n)$, we understand the product of the effective
power of transmitter $i$ and of the random fading from this
node to receiver $y_j$. The random (vector valued) processes
$\{F_i^j(n):n\}$ are assumed to be i.i.d. in $i$ and $j$ given
$\{X_i,y_i\}$, and we denote by $F$ the
generic marginal random variable $F_i^j(n)$.
% and the components $(F_i^j,j)$
%are assumed to be identically distributed (distributed as a
%generic random variable (r.v.) 
%denoted by $F$) with mean $1/\mu$ assumed finite. 
%The components might be dependent, in particular due to the 
%common effective transmission power of node $X_i$, when this
%power is not constant for all nodes.
In the case of constant effective transmission power $1/\mu$ and 
Rayleigh fading (which is our default scenario), 
$F$ is exponential with mean $1/\mu$ 
(see e.g.~\cite[p.~50 and~501]{TseVis2005}). However,  one also consider also non exponential cases
to analyze other types of fading such as e.g. Rician
or Nakagami scenarios (cf~\cite[Sec.~23.2.4]{FnT2})
or simply the case without fading (when 
$F\equiv 1/\mu$ is deterministic).
Note that we do not have specified yet the dependence between the
virtual powers $F_i^j(n)$ for different time slots $n$. This is
done in what follows.
\begin{itemize}
\item By the  {\em fast fading} case we understand 
the scenario when  the random variables $F_i^j(n)$
are i.i.d. in~$n$ (recall, that the default option is that they are
also i.i.d. in $i,j$);~%
~\footnote{%
Note that our fast fading means that it remains constant over a slot
duration and can be seen as i.i.d. over different time slots.
This might not correspond to the terminology used in many papers of literature,
where fast fading means that the channel conditions fluctuate much over a given time slot.}
\item The {\em slow fading} case is that where $F_i^j(n)\equiv F_i^j$, 
for all $n$.
\end{itemize}

\item\label{i.W} $W_i(n)$ represents the thermal noise at the receiver $y_i$ 
at time $n$. The processes $\{W_i(n):n \}$ are  independent 
in~$i$ given $\{y_i\}$, with the  generic marginal random variable
denoted by $W$. For the time dependence, 
 one can consider both  $W_i(n)$ independent in $n$ 
({\em fast noise}) and constant $W_i(n)\equiv W_i$ ({\em slow noise}).
\end{itemize}

%Note that the point process of transmitters 
%$\Phi^1(n)=\sum_i \delta_{X_i}\ind(e_i(n)=1)$ varies over time due to
%the MAC decisions.
%So does the interference process  at the receiver of the node $X_i$:
% $I_i^1(n)=\sum_{X_j\in\widetilde\Phi^1(n),\, j \ne i
%}F_j^i(n)/l(|X_j-y_i(n)|)$ that, mathematically, is an example of the
%{\em Shot-Noise (SN)}  
%of $\widetilde\Phi^1(n)\setminus\{X_i\}$.

\subsection{Receiver Models}
\label{ss.receiver}
We first recall the simple receiver model used in~\cite{jsac} and then
propose a few possible extensions.

\subsubsection{Bipolar Receiver Model} In this model the receivers 
are {\em external to the MANET p.p.} $\Phi=\{X_i\}$. We assume that, given $\Phi$, 
the random variables $\{X_i-y_i\}$ are i.i.d random vectors with
$|X_i-y_i|=r$; i.e. each {\em receiver is at distance} $r$ from its transmitter.
Note that the receivers of different MANET nodes are different (almost surely).
% and that they do not vary in time $y_i(n)=y_i$.~\footnote{%
%In fact our analysis of the local delays in this time does not depend
%on whether $y_i$ are static or re-sampled at each time slot within
%the same distance~$r$.}

\subsubsection{Nearest Receiver  models}
In practice, some routing algorithm
specifies the receivers(s) (relay node(s)) of each given transmitter. 
In what follows focus on two {\em nearest receiver  models} 
where each transmitter selects its receiver as close by as possible
is some set of {\em potential receivers} $\Phi_0$, common to all MANET nodes
$\Phi$; i.e.  $y_i=Y_i^*=\argmin_{Y_i\in\Phi_0, Y_i\not=X_i}\{|Y_i-X_i|\}$.
Two incarnations are considered 
\paragraph{Independent Poisson Nearest Receiver (IPNR) Model}
In this model we assume that the potential receivers form some
stationary P.p.p. $\Phi_0$, of intensity $\lambda_0$,
 which is {\em independent} of  (and in
particular {\em external} to)  the MANET $\Phi$. 
\paragraph{MANET Nearest Neighbor (MNN) Model}
All the nodes of the MANET are considered as potential
receivers; i.e. $\Phi_0=\Phi$.~%
\footnote{Both IPNR and MNN models require some additional
  specifications on what 
happens if two or more transmitters pick the same receiver
and, the MNN model,  what
happens if the picked receiver is also transmitting.
Our analysis applies to the situation when the SINR threshold
$T>1$ (cf.~\ref{ss.SINR}), which excludes multiple receptions by
a given receiver and  
simultaneous emission and reception.}

\subsection{Mean Path-loss Model}
\label{ss.OPL}
Below, we assume that the receiver 
of node $i$ receives a power from the transmitter located at 
node $j$ at time $n$ which is equal to $F_i^j(n)/l(|X_j-y_i|)$,
where \hbox{$|\cdot|$} denotes
the Euclidean distance on the plane and $l(\cdot)$
is the path loss function.
An important special case consists in taking 
\begin{equation}\label{simpl.att}
l(u)=(Au)^{\beta} \quad \text{for $A>0$ and $\beta>2$.}
\end{equation} 
Other possible choices of path-loss function avoiding the
pole at $u=0$ consist in taking e.g. $\max(1,l(u))$,
$l(u+1)$, or  $l(\max(u,u_0))$.

\subsection{SINR Coverage}
\label{ss.SINR}
In this paper we mainly focus on the SINR coverage/outage scenario:
we will say that transmitter $\{X_i\}$ 
{\em covers} (or {\em is successfully received by}) its receiver
$y_i$ at time slot $n$ if
\begin{equation}\label{eq:SINR}
\mbox{SINR}_i(n)=\frac{F_i^i(n)/l(|X_i-y_i|)}{W_i(n)+I_i^1(n)}\ge T\,,
\end{equation}
where $I_i^1(n)=\sum_{X_j\in\widetilde\Phi^1(n),\, j\not=i}
F_j^i(n)/l(|X_j-y_i|)$ 
is the {\em interference} at receiver $y_i$ at time $n$; 
i.e., the sum of the signal powers 
received by $y_i$ at time $n$ from  all the nodes in 
$\Phi^1(n)=\{X_j\in\Phi: e_j(n)=1\}$ except $X_i$.
In mathematical terms, $I_i^1(n)$ is 
an instance of {\em Shot-Noise (SN)} field generated by 
of $\widetilde\Phi^1(n)\setminus\{X_i\}$.

Denote by $\delta_i(n)$ the indicator that~(\ref{eq:SINR}) holds,
namely, that location $y_i$ is covered by transmitter $X_i$ with
the required quality at time $n$.

%The scenarios described in \S~\ref{s.beyond.dipole}
%where a snapshot of the network is made of two independent Poisson point 
%processes of transmitters and receivers, are compatible with
%different time-space scenarios:
%\begin{itemize}
%\item that where the p.p. $\Phi^0$ of receivers is fixed once and for all
%and independent of the sequence $\{\Phi^1(n)\}$ defined above generalizes
%the independent receiver model;
%\index{independent receiver model}
%\item that where the p.p. of receivers at time $n$ is $\Phi^0(n)= \Phi-\Phi^1(n)$
%generalizes the MANET receiver model.
%This last model will be the basis of Chapter \ref{ch:OpporR},
%\index{MANET receiver model}
%\end{itemize}
%The default option for the selection of the receivers will be nearest neighbor routing.

\subsection{Typical MANET node}
\label{ss.Palm}
Let $\Pro^0$ denote the Palm
distribution of the P.p.p. $\Phi$ (cf~\cite[Sec.~10.2.2]{FnT1}).
Under this distribution, 
the MANET nodes are located at $\Phi\cup\{X_0=0\}$,
where $\Phi$ is a copy of the stationary P.p.p.
(cf Slivnyak's theorem; ~\cite[Th.~1.4.5]{SKM95}).
Under $\Pro^0$, the other random objects/marks of the model, $(e_i(n),\bfF_i(n))$ and $W_i(n)$ as well as
$y_i(n)$ in the Bipolar receiver model, are i.i.d. given
$\Phi\cup\{X_0=0\}$, and have the same law as their original distribution.
(For more details on Palm theory  cf. e.g.~\cite[Sections~1.4, 2.1
  and 10.2]{FnT1}.) 
In the IPNR model, the potential receiver p.p. $\Phi_0$ remains independent of
$\Phi\cup\{X_0=0\}$ and Poisson-distributed; in the MNN model the receiver p.p.
is still determined by the MANET configuration $\Phi\cup\{X_0=0\}$ (with the receiver
of a node being its nearest neighbor).
The node $X_0=0$, considered under $\Pro^0$, is called {\em the typical MANET node}.

\section{Local Delay}
\resetcounters
\label{s.TheLD}
The {\em local delay of the typical node
is the number of time slots needed for node $X_0=0$
(considered under the Palm probability $\Pro^0$ with respect to $\Phi$)
to successfully transmit}: 
$$\LD =\LD _0=\inf\{n\ge1: \delta_0(n)=1\}\,.$$ 
This random variable depends on the origin of time (here 1) but
we focus on its law below, which does not depend on the chosen
time origin.

The main objective of this paper is to study~$\E^0[\LD ]$ under the
full separation of time scales described in the introduction.  
Let  $\calS$ denote all the {\em static elements of the network model}: 
i.e. the elements which are random but which do not vary with time $n$. 
In all models, we have all locations $\Phi,\{y_i\} \in\calS$. Moreover, in the
slow fading model, we have $\{\bfF_i(n)=\bfF_i\}\in\calS$ and similarly in
the slow noise model, $\{W_i(n)=W_i\}\in\calS$. 

Given a realization of all the elements of $\calS$, denote by 
\begin{equation}
\label{chalopic}
\pi_c(\calS)=\E^0[e_0(1)\delta_0(1)\,|\,\calS] 
\end{equation}
the conditional probability, given $\calS$, that $X_0$
is authorized by the MAC to transmit and that this transmission is successful 
at time $n=1$.
Note that due to our time-homogeneity, this conditional probability
does not depend on $n$. The following result allows us to express $\E^0[\LD ]$.
\begin{Lemma}\label{l.delay-basic}
We have
\begin{equation}\label{e.delay-basic}
\E^0[\LD _0]
%=\E^0\Bigl[\frac1{\E^0[e_0(1)\delta_0(1)\,|\,\calS]}\Bigr]
=\E^0\Bigl[\frac1{\pi_c(\calS)}\Bigr]\,.
\end{equation}
\end{Lemma}

\rem
One can interpret  $\pi_c(\calS)$ as the {\em (temporal) rate of
successful packet transmissions} (or the {\em throughput}) 
of node $X_0$ given all the static elements of the network.
Its inverse $1/\pi_c(\calS)$ is the local delay of this node in this environment.
In many cases, this throughput is a.s. positive (so that we will have
a.s. finite delays)  
for all static environments. If this last condition holds true, by
Campbell's formula (cf~\cite[Eq.~(10.14)]{FnT1}),
almost surely, all the nodes have
finite mean delays and  positive throughputs. However, the spatial
irregularities of the network imply that this throughput varies from
node to node, and in a Poisson configuration, one can find
nodes which have an arbitrarily small throughput (and consequently an arbitrarily large
delay). The mean local delay of the typical node 
 $\E^0[\LD ]$ is the spatial average 
of these individual mean local delays. A finite mean indicates that the
fraction of nodes in bad shape (for throughput or delay)
is in some sense not significant. In contrast, $\E^0[\LD ]=\infty$ 
indicates that an important fraction of the nodes are in
a bad shape. This is why the finiteness of the mean local delay of the
typical node is an
important indicator of a good performance of the network.  

\begin{proof}(of Lemma~\ref{l.delay-basic})
Since the elements that are not in $\calS$  change over time in the
i.i.d. manner given a realization of the elements of $\calS$, 
the successive attempts of node $X_0$ to access to
the channel and successfully transmit at time $n\ge 1$ are independent (Bernoulli)
trials with probability of success $\pi_c(\calS)$. 
The local delay $\LD =\LD _0$ is then a geometric random variable (the number of trials
until the first success in the sequence of Bernoulli trials) with parameter $\pi_c(\calS)$.
Its (conditional) expectation (given $\calS$) is known to be 
$\E^0[\LD \,|\,\calS]=1/\pi_c(\calS)$.
The result follows by integration with respect to the
distribution of  $\calS$.
\end{proof}

\begin{Exe}\label{exe.basic}
In order to understand the reasons for which $\E^0[\LD ]$ may or
may not be finite, consider the following two extremal situations.
Suppose first that the whole network is independently  re-sampled 
at each time slot (including node locations
$\Phi$, which is {\em not} our default option).
Then $\calS$ is empty (the $\sigma$-algebra generated by it is
trivial) and the temporal rate of successful transmissions is equal to 
the space-time average rate $\pi_c(\calS)=\E^0[e_0(1)\delta_0(1)]=p\,p_c$,
where $p_c$ is the space-time probability of  coverage 
for the typical node in  the corresponding 
receiver model. Consequently, in this case of extreme variability (w.r.t. time),
we have $\E^0[\LD ]=1/p_c<\infty$  provided $p_c>0$, 
which holds true under very mild assumptions, e.g. for all considered
receiver and path-loss models with fast or slow Rayleigh fading model
and any noise model  provided, $0<p<1$ (cf.~\cite{jsac} for the
evaluation of $p_c$ for the Poisson 
Bipolar model).

On the other hand, if nothing varies over time (including MAC status,
which again is ruled out in our general assumptions), we 
have $\pi_c(\calS)=e_0(1)\delta_0(1)$ (because the conditioning on
$\calS$ determines $e_0(1)\delta_0(1)$ in this case).
In this case under very mild 
assumptions (e.g. if $p<1$), this temporal rate
$e_0(1)\delta_0(1)$ is zero with positive probability, making $\E^0[\LD ]=\infty$.
Note that in this last case, some nodes in the MANET
succeed in transmitting packets every time slot, whereas
others never succeed.
Having seen the above two extremal cases, it is not difficult to 
understand that the mean local delay of the typical node very much depends on  
how much the time-variability ``averages out'' the spatial
irregularities of the distribution of nodes in the MANET.
\end{Exe}

Note that by Jensen's inequality,
$$\E^0[\LD ]\ge\frac1{\E^0[\pi_c(\calS)]}=\frac1{p_c}\,.$$
The inequality is in general strict and we may have
\hbox{$\E^0[\LD ]=\infty$} while $p_c>0$.

In the remaining part of this section we will study several particular
instances of space-time scenarios.

\subsection{Local Delays in Poisson Bipolar Model}
In the Poisson bipolar model, we assume a static repartition for the MANET
nodes $\Phi$ and for their receivers $\{y_i\}$. The MAC variables
$e_i(n)$ are i.i.d. in $i$ and $n$. All other elements
(fading and noise) have different time-scenarios.

\subsubsection{Slow Fading and Noise Case}
Let us first consider the situation where $\{\bfF_i\}$ and $W$ are static.
\begin{Prop}
Assume the Poisson Bipolar network model with slow fading and slow noise. 
If the distribution of $F,W$ is such that 
$\Pr\{\,WTl(r)>F\,\}>0$, then $\E^0[\LD ]=\infty$.
%\begin{itemize}
%\item If the distribution of $F,W$ is such that 
%$\Pr\{\,WTl(r)>F\,\}>0$ then $\E^0[L]=\infty$.
%\item If  $W=0$ a.s.  then $\E^0[L]<\infty$.{\bf ????}
%\end{itemize}
\end{Prop}
\begin{proof}
%Assume $\Pr\{\,WTl(r)>F\,\}>0$.
We have
\begin{eqnarray*}
\pi_c(\calS)&=&p\E^0[e_0(1)\delta_0(1)\,|\,
\calS]\\
&=&p\Pr^0\{F^0_0\ge Tl(r)(W_0+I_0^1(0))\,|\,\calS\}\\
&\le&p\ind(F^0_0\ge Tl(r)W)\,.
\end{eqnarray*}
The last indicator is equal to 0 with non-null probability
for our assumptions. Using~(\ref{e.delay-basic}), we conclude that 
$\E^0[\LD ]=\infty$.
%Assume now $W=0$ (we will call such scenario interference-limited
%case. Fix $\Phi$, the location of the receiver $y_0$ 
%and $F_i^j$ for all $i,j$. 
%
%{\bf In fact, I do not know how to prove it. Is it true?}
%Since by our common assumption on the path-loss function 
%we can assume that the  shot noise generated by 
% all the nodes $I=\sum_jF_j^j/l(|X_j-y_0|)$ is finite for our
% fixed realizations  of $\Phi, y,F_i^j$ ($i,j$).
% We can find thus a radius $R<\infty$ (which my depend on such that 
%$I(R)=\sum_j\ind(|X_j>R)F_j^j/l(|X_j-y_0|)$
\end{proof}

\subsubsection{Fast Fading Case}
The following auxiliary result is useful when studying fast
Rayleigh fading.
\begin{Lemma}\label{l.SN-LT-condtional-inverse}
Consider the Poisson shot-noise $I=\sum_{X_i\in\Phi}G_i/l(|X_i|)$,
where $\Phi$ is some homogeneous Poisson p.p. with
intensity $\alpha$ on $\ir^2$, 
$G_i$ are i.i.d. random variables with Laplace transform
$\calL_G(\xi)$ and $l(r)$ is any response function (in our case it
is always some path-loss function). 
Denote by $\calL_I(\xi\,|\,\Phi)=\E[e^{-\xi I}\,|\,\Phi]$ the
conditional Laplace transform of $I$ given $\Phi$. Then
$$\E\Bigl[\frac1{\calL_{I}(\xi\,|\,\Phi)}\Bigr]=
\exp\biggl\{-2\pi\alpha\int_0^\infty
v\Bigl(1-\frac1{\calL_G(\xi/l(v))}\Bigr)\,\md v\biggr\}\,.$$
\end{Lemma}
\begin{proof}
By the independence of $G_i$ given $\Phi$, we have 
\begin{eqnarray*}
\E[e^{-\xi I}\,|\,\Phi]&=&
\prod_{X_i\in\Phi}\E[e^{-\xi\calL_G(\xi/l(|X_i|))}]\\
&=&\prod_{X_i\in\Phi}\calL_G(\xi/l(|X_i|))\\
&=&\exp\Bigl\{\sum_{X_i\in\Phi}\log(\calL_G(\xi/l(|X_i|)))\Bigr\}\,.
\end{eqnarray*}
Taking the inverse of the last expression and using
the known formula for the Laplace transform of the Poisson p.p. 
(it can be derived
from the formula for the Laplace functional of the Poisson p.p.;
see e.g.~\cite{daley},  and was already used e.g.~\cite{jsac}).
we obtain 
\begin{equation}\label{e.SN-LT-condtional-inverse-nonhomo}
\E\Bigl[\frac1{\calL_{I}(\xi\,|\,\Phi)}\Bigr]=
\exp\biggl\{-\alpha\int_{\ir^2}
\Bigl(1-e^{-\log(\calL_G(\xi/l(|x|)))}\Bigr)\,\md x\biggr\}\,.
\end{equation}
Passing to polar coordinates completes the proof.
\end{proof}
%\rem The extension of the above result to a non-homogeneous Poisson
%p.p. $\Phi$ is straightforward
%and consists in replacing the integral
%$\alpha\int_{\ir^2}(\ldots)\,\md x$ with respect to the Lebesgue
%measure $\alpha\,\md x$ on $\ir^2$ 
%in~(\ref{e.SN-LT-condtional-inverse-nonhomo}) above, 
%by the integral of the same function with respect to the intensity
%measure of the considered non-homogeneous Poisson p.p.
%This extension will be useful when we will consider the local delay 
%in the MNN model in Section~\ref{ss.LD-NRM}, where the conditioning on the
%distance to the receiver modifies the intensity of the process of
%interferrers (cf.  Proposition~\ref{p.pc-PoNR-MNN})
%or in the MNN model with closest receiver in a cone (see
%Remark~\ref{rem:alo-cone} below).

Coming back to local delays,
let us consider now  the situation where the random variables $\{\bfF_i(n)\}$
are i.i.d. in $n$. We consider only the Rayleigh fading case.
\begin{Prop}\label{p.delay-bipolar-fast-fading}
Assume the Poisson Bipolar network model with fast Rayleigh fading. 
In the case of fast thermal noise, we have
$$\E^0[\LD]=
\frac1{p}
%\calL_W(\mu Tl(r))}
\calD_W(Tl(r))
\exp\Bigl\{2\pi p\lambda\int_0^\infty
\frac{vTl(r)}{l(v)+(1-p)Tl(r)}\,\md v\Bigr\}\,,
$$
where 
\begin{itemize}
\item $\calD_W(s)=\calD_W^{slow}(s)=\calL_W(-s)$ for the slow noise
  case,
\item $\calD_W(s)=\calD_W^{fast}(s)=1/\calL_W(s)$ for the fast noise
  case.
\end{itemize}
%In the case of slow noise we have
%$$\E^0[\LD ]=
%\frac1{p}\calL_W(-\mu Tl(r))
%\exp\biggl\{2\pi p\lambda\int_0^\infty
%\frac{vTl(r)}{l(v)+(1-p)Tl(r)}\,\md v\,\Bigr\}\,.
%$$
\end{Prop}
\begin{proof}
In the fast Rayleigh fading case, we have
$\pi_c(\calS)=\Pr\{F\ge Tl(r)(W+I^1)\,|\,\Phi\}$
for the fast noise case and $\pi_c(\calS)=\Pr\{F\ge
Tl(r)(W+I^1)\,|\,\Phi,W\}$
for the slow noise model.
Using the assumption on $F$, we obtain 
$$\pi_c(\calS)=\calL_W(\mu Tl(r))\E[e^{-\mu Tl(r) I^1}\,|\,\Phi]$$
in the fast noise case
and 
$$\pi_c(\calS)=e^{-\mu WTl(r)}\E[e^{-\mu Tl(r) I^1}\,|\,\Phi]$$
for the slow noise case.
The result then follows from~(\ref{e.delay-basic}) 
and Lemma~\ref{l.SN-LT-condtional-inverse}
with $G=eF$. Note that in this case
$\calL_G(\xi)=1-p+p\calL_F(\xi)$, which gives
$\calL_{eF}(\xi)=1-p+p\mu/(\mu+\xi)$.
\end{proof} 

\subsection{Local Delay in the Nearest Receiver Models}
\label{ss.LD-NRM}
In this section we study the IPNR and the MNN receiver models.
We work out formulas for the mean local delay of the typical node 
under the following conditions: {\em fast Rayleigh fading}
and fast or slow noise of arbitrary distribution.

\begin{Prop}
\label{p:delay-NN-main}
Assume fast Rayleigh fading.
\begin{itemize}
\item In the IPNR model, we have 
\begin{equation}\label{e.delay-INR-main}
\!\!\!\!\E^0[\LD ]\!=\!\frac{2\pi\lambda_0}{p}\!\int_0^\infty \!\!\!re^{-\pi\lambda_0r^2}
\calD_W(\mu Tl(r))\,\calD^{INR}_I(\mu Tl(r))\md r 
\end{equation}
where 
\begin{equation}\label{eq:ch9basicld}
\calD_{I}^{INR}(s)=
\exp\left\{2\pi\lambda \int_0^\infty \frac{ps}{l(v)+ (1-p)s} v\, \md
  v\right\}
\end{equation}
and $\calD_W(s)$ is as in
Proposition~\ref{p.delay-bipolar-fast-fading}.
\item In the MNN model, we have 
\begin{eqnarray}\label{e.delay-MNN-main}
\lefteqn{\E^0[\LD ]=\frac{2\pi\lambda}{p(1-p)}}\\[0.5ex]
&&\times\int_0^\infty re^{-\pi\lambda r^2}
\calD_W(\mu Tl(r))\,\calD^{MNN}_I(r,\mu Tl(r))\,\,\md r\,, \nonumber
\end{eqnarray}
where 
\begin{eqnarray}\label{eq:ch9basiclddsm}
\lefteqn{\calD_{I}^{MNN}(r,s)=\exp \biggl\{
\lambda\pi \int_0^\infty \frac{ps}{l(v)+ (1-p)s} v\,\md v
}\nonumber\\
&&\hspace{-3em}+\lambda  
\int_{\theta=-\frac \pi 2}^{\frac \pi 2} 
\int_{v>2r \cos \theta }
\frac{ps}{l(v)+ (1-p)s} v\, \md v \md \theta
\biggr\}
\end{eqnarray}
and $\calD_I(s)$ is as above (as in
Proposition~\ref{p.delay-bipolar-fast-fading}).
\end{itemize}
\end{Prop}
\begin{proof}
We condition on the location of the nearest
neighbor $y_0=Y^*_0$ of $X_0=0$ under $\Pro^0$. 
In the IPNR model the  distance from the origin to $Y_0^*$ is known to
have the following distribution (under both the Palm and the stationary law):
$\Pro\{|Y_0^*|>r\}=e^{-\pi\lambda_0 r^2}$. Due to the
independence assumption, in this model, given the location of the receiver $Y_0^*$,
the distribution of the MANET nodes $\Phi$ remains unchanged under $\Pro^0$.
The remaining part of the proof follows the same lines as that of
Proposition~\ref{p.delay-bipolar-fast-fading}.

For the MNN model, recall from Section~\ref{ss.Palm} that under 
$\Pro^0$, the nodes of $\Phi\setminus\{X_0\}$
%, in which the receiver $y_0$ is taken as the
%nearest node $Y_0^*$ of the origin,
are distributed as those of the homogeneous Poisson p.p.
Thus the distance $|Y^*_0-X_0|=|Y^*_0|$ has the same distribution as in
the IPNR model with $\lambda_0=\lambda$.
However, in the MNN model, given some particular location of $y_0=Y^*_0$,
one has to take the following fact into account: there are
{\em no MANET nodes} (thus, in particular, no interferers) 
in $B_0(|y_0|)$. Consequently, under $\Pro^0$, given $Y_0^*=y_0$, 
the SN $I^1_0$ in~(\ref{eq:SINR}) is no longer
driven  by the stationary Poisson p.p. of intensity $\lambda_1$,  
but as the SN of $\Phi^1$ given that there are no nodes of $\Phi$ in
$B_0(|y_0|)$. Note that the location $y_0$ at which we evaluate this
last SN is on the boundary (and not in the center)
of the empty ball. By the strong Markov property of Poisson p.p.
(cf.~\cite{Zuyev2006}), the
distribution of a Poisson p.p. given that $B_0(|y_0|)$ is empty
is equal to the distribution of the (non-homogeneous)
Poisson p.p. with intensity equal to 0 in $B_0(|y_0|)$ and 
$\lambda_1$ outside this ball. Putting these arguments together, 
and exploiting the rotation invariance of the picture conclude the proof.
\end{proof}

Notice that the integrals in (\ref{eq:ch9basicld}) 
and (\ref{eq:ch9basiclddsm}) are finite for any of the path-loss
models suggested in Section~\ref{ss.OPL}.
However, the  outer integrals (in $r$) in~(\ref{e.delay-INR-main})
and~(\ref{e.delay-MNN-main}) may be infinite. In order to study this
problem note first that we have the following bounds in the MNN model:

\begin{Remark}
\label{remal:dsmbounds}
In the MANET receiver case, we have the bounds
\begin{equation}
\label{eq:ch9basiclddsm-bound}
\Bigl(\calD_{I}^{INR}(s)\Bigr)^{1/2}
 \le \calD_I^{MNN}(r,s)\le 
\calD_{I}^{INR}(s)\,.
\end{equation}
\end{Remark}

\section{Wireless contention phase transition for the local
  delay}
\label{s.Phase-Transition}
In this section we will show that under quite natural assumptions
them mean local delay of the typical node  can be infinite.
For some models, it can exhibit the following phase transition:
$\E^0[\LD ]<\infty$ or $\E^0[\LD ]=\infty$ depending on the
model parameters (as $p$, distance $r$ to the receiver, or the mean fading $1/\mu$).

\subsection{Bipolar Model}
We begin with the simple Bipolar receiver model.
Proposition~\ref{p.delay-bipolar-fast-fading} shows
that in the fast fading and noise case, $\E^0[\LD ]<\infty$;
indeed, $\int_0^\infty v/l(v)\,\md v<\infty$.
However for the fast fading, slow noise case 
the finiteness of the mean local delay of the typical node
depends on whether $W$ has finite
{\em exponential moments} of order $Tl(r)\mu$. 
This is a rather strong assumption concerning the tail distribution
function of $W$. Often this moment is finite only for 
some sufficiently small value of $Tl(r)\mu$.

To see the wireless contention phase transition in this model consider
the following example.
\begin{Exe}
\label{R:wireless-contention-phase-trans-bipolar}
Let us assume exponential noise with mean $1/\nu$.
Then $\calL_W(-\xi)=\nu/(\nu-\xi)<\infty$ provided 
$Tl(r)\mu<\nu$ and infinite for $Tl(r)\mu>\nu$.
This means that in the corresponding Poisson Bipolar MANET 
with a Rayleigh fading, exponential noise, 
the mean local delay 
of the typical node is finite whenever $Tl(r)<\nu/\mu$ and infinite otherwise.
Here are a few incarnations of this phase transition:
\begin{itemize}
\item For fixed mean transmission power $\mu^{-1}$ (we recall that a typical
situation is that where fading has mean 1 and where $\mu^{-1}$ is actually
the effective transmission power) and mean
thermal noise $\nu^{-1}$, there is a threshold
on the distance $r$ between transmitter and receiver below which
mean local delay of the typical node is finite and above which they
are infinite; 
\item For fixed mean thermal noise $\nu^{-1}$ and fixed distance $r$,
there is a threshold on mean transmission power $\mu^{-1}$ {\em above}
which the mean local delay of the typical node  is finite and {\em
  below} which it is  infinite; 
\item For fixed mean transmission power $\mu^{-1}$ and fixed distance $r$,
there is a threshold on mean thermal noise power $\nu^{-1}$ {\em below}
which the mean local delay of the typical node 
is finite and {\em above} which it is infinite.
\end{itemize}
The fact that all transmissions contend for the shared wireless channel may lead
to infinite mean local delays of the typical node
if the system is stressed by either of the
phenomena listed above: too distant links, a too high thermal noise or a too 
transmission power.
\end{Exe}
 
\begin{Remark}{\em (RESTART)}\label{R.restart}
There is a direct interpretation of the 
local delay of the typical node in terms of the so called {\em RESTART
  algorithm}:% 
\index{RESTART algorithm}
assume a file of random size $B$ is to be transmitted over
an error prone channel. Let $\{A_n\}_{n\ge1}$ be the sequence
of channel inter-failure times.
If $A_1>B$ (resp. $A_1\le B$), the transmission succeeds
(resp. fails) at the first attempt. 
If the transmission fails at the first attempt, one has
to restart the whole file transmission in the second attempt and so on. Let 
$$ N= \inf \{n\ge 1 \ s.t.\ A_n>B\}$$ 
be the first attempt where the file is successfully transmitted.
In the classical RESTART scheme, the sequence $\{A_n\}_{n>0}$
is assumed to be i.i.d. and independent of $B$.
It can then be proved (see \cite{Asmussen}) that when $B$
has infinite support and $A_n$ is light tailed (say exponential), then $N$
is heavy tailed. This observation comes as a surprise because one
can get heavy tails (including infinite first moments) in situations
where $B$ and $A_n$ are both light tailed.

Consider the fast fading, slow noise case (and ignore the
interference for simplicity)
Then the local delay of the typical node  can be seen as an instance of this
algorithm with the following identification:
$A_n=F_0^0(n)e_0(n)$ and $B=TW l(r)$.
In the next section, 
we will see other incarnation of the above RESTART algorithm 
in the nearest-receiver models with
deterministic $W$, where the role of the unboundedness of $B$ is
played by the distance to the receiver;
cf.~Remark~\ref{R.restart-cont}.

The above interpretation in terms of the RESTART scheme formally shows 
that the local delay of the typical node is heavy tailed 
and thus it is not surprising that in certain cases its mean is infinite.
However, the physical phenomena at hand are quite different here and
in the classical RESTART context, so
let us now comment on what exactly this heavy tailedness means in our
MANET context. 
Let $B_0(R)$ be the ball centered at $0$ of  radius $R$ and let
$\LD_i$  be the number of time slots required by the node $X_i$ to
transmit a packet.
The ergodic interpretation of the Palm probability
implies that, for all $m$,
$$\Pro^0\{\,\LD_0>m\,\} =\lim_{R\to \infty} \frac 1 {\Phi(B_0(R))} 
\sum_{X_i\in B_0(R)}\ind(\LD_i>m),$$
where the last limit is in the almost sure sense.
The fact that the distribution of $\LD_0$ is heavy tailed under
$\Pro^0$ means that the (discrete) law $\Pro^0\{\,\LD_0>m\,\}$ 
has no exponential moments. In view of the above ergodic
interpretation, this is equivalent to saying that the asymptotic
fraction of MANET modes which experience a local delay
of more than $m$ time slots decreases slowly with $m$ (more slowly than any
exponential function).

Finally,  we remark  that the fact that the {\em mean local delay of
  the typical node 
is infinite does not imply that the mean throughput of the typical node
is null}.  The last quantity boils down to the probability
of success of node $X_0$ under $\Pro^0$, i.e., to $p_c$ and,
as already mentioned in Example~\ref{exe.basic},  it is positive for all
the considered models.

\end{Remark}

\subsection{Nearest Receiver Models}
In order to analyze mean local delays in these  more complex receiver
models will study separately the impact of the thermal noise and
of the interference.
\subsubsection{Noise Limited Networks}
\label{sss:delay-noise-limitted-NN}
Consider first the IPNR and MNN models under the assumption that  
the interference is perfectly canceled
(and that only noise has to be taken into account).
In what follows we consider the fast noise scenario.

\paragraph{IPNR model}
The  following result follows from 
(\ref{e.delay-INR-main}) with $\calD_I(s)=1$
and with $\calD_W$ given in
Proposition~\ref{p.delay-bipolar-fast-fading}:
\begin{Cor}
In the IPNR model with fast Rayleigh fading and fast noise,
if interference is perfectly canceled, then 
\begin{equation*}
\E^0[\LD] = 2\pi \lambda_0 \int_0^\infty
\frac {r \exp(-\pi \lambda_0 r^2)} {p {\cal L}_W(\mu l(r)T)} \md r.
\end{equation*}
\end{Cor}
Hence, for the simplified path loss function~(\ref{simpl.att})
$\E^0[\LD]<\infty$ whenever
\begin{equation}
\label{eq:oppsufcondnoisenndir}
{\cal L}_W(\xi) \ge 
\eta \exp\left\{-\pi \lambda_0 \left(\frac{\xi}{ \mu T A^{\beta}}\right)^{2/\beta}\right\}
\left(\xi^{2(1+\epsilon)/\beta}\right),
\  \xi\to \infty,
\end{equation}
for some positive constants $\epsilon$ and $\eta$,
and whenever some natural local integrability conditions also hold.
This condition requires that there be a sufficient probability mass of $W$
in the neighborhood of 0. For instance,
under any of the path-loss models suggested in Section~\ref{ss.OPL}, this holds true for a thermal
noise with a rational Laplace transform (e.g. Rayleigh).

The condition~(\ref{eq:oppsufcondnoisenndir}) is sharp in the sense
that when ${\cal L}_W(\xi)$ is 
asymptotically smaller that the expression in the right-hand-side
of~(\ref{eq:oppsufcondnoisenndir}) with $(1+\epsilon)$ replaced by
$(1-\epsilon)$  
%\begin{equation}
%\label{eq:oppsufcondnoisennrec}
%{\cal L}_W(\xi) \le 
%\eta \exp\left\{-\pi \lambda_0 \left(\frac{\xi}{ \mu T A^{\beta}}\right)^{2/\beta}\right\}
%\left(\xi^{2(1-\epsilon)/\beta}\right),
%\  x\to \infty,
%\end{equation} 
for some positive constants $\epsilon$ and $\eta$,
then $\E^0[\LD]=\infty$. This is the case, e.g. when $W$ is a
positive constant.

\paragraph{MNN model}
In the MNN  model (with $0<p<1$) and the simplified path loss
function~(\ref{simpl.att}), similar arguments show that 
the same threshold as above holds with 
\begin{equation}
\label{eq:oppsufcondnoisennds}
{\cal L}_W(\xi) \ge 
\eta \exp\left\{-\pi \lambda \left(\frac{\xi}{ \mu T A^{\beta}}\right)^{2/\beta}\right\}
\left(\xi^{2(1+\epsilon)/\beta}\right),
\ \xi\to \infty,
\end{equation}
implying that $\E^0[\LD]<\infty$ and a similar converse statement.

\subsubsection{Interference Limited Networks}
\label{sss:delay-interf-limitted-NN}
\index{network!interference limited}
In this section we assume  and $W\equiv 0$. 
\paragraph{IPNR model}
In the IPNR  case with the simplified path loss
function~(\ref{simpl.att}), 
using the fact that
\begin{equation}\label{e.PINR-interf}
2 \pi \int_0^\infty \frac{pT l(r)}{l(v)+ (1-p)T l(r)} v\, \md v = p 
(1-p)^{\frac 2 \beta-1} T^{\frac 2 \beta} K(\beta) r^2, 
\end{equation}
with 
\begin{equation}
\label{eq:kdebeta}
K(\beta)=\frac{2\pi\Gamma(2/\beta)\Gamma(1-2/\beta)}{\beta}=
\frac{2\pi^2}{\beta\sin(2\pi/\beta)}\, .
\end{equation}
Using the above observations we  get the following result
from~(\ref{e.delay-INR-main}) 
and~(\ref{eq:ch9basicld}): 
\begin{Cor}\label{c:PINR-delay-interflim}
In the IPNR model with $W=0$, fast Rayleigh fading and the 
path loss function~(\ref{simpl.att}), we have
\begin{equation*}
\E^0[\LD] = 2\pi \lambda_0 \frac 1 p 
\int_0^\infty r \exp\left(-\pi \lambda_0 r^2 + \lambda \theta(p,T,\beta) r^2\right) \md r\, ,
\end{equation*}
with
\begin{equation}\label{e:theta-delay}
\theta(p,T,\beta)= \frac p {(1-p)^{1 -\frac 2 \beta}}
T^{\frac 2 \beta} K(\beta) \,.
\end{equation}
\end{Cor}
Notice that $\theta(p,T,\beta)$  is increasing in $p$ and in $T$.
We hence get the following {\em incarnation of the wireless contention phase transition}:
\begin{itemize}
\item If $p\ne 0$ and  $\lambda_0 \pi >  \lambda \theta(p,T,\beta)$, then 
\begin{eqnarray*}
\E^0[\LD]&=&
\frac 1 p 
\frac{\pi \lambda_0 }{\pi \lambda_0-  \lambda \theta(p,T,\beta)} \\
&=&\frac 1 p 
\frac{\lambda_0 }{\lambda_0- \lambda 
\frac 2 \beta \Gamma(\frac 2 \beta)\Gamma(1-\frac 2 \beta)
p(1-p)^{2/\beta-1} T^{2/\beta}}\\
&<& \infty\, .
\end{eqnarray*}
\item If either $\lambda_0 \pi  <  \lambda \theta(p,T,\beta)$ or $p=0$, 
then $\E^0[\LD] =\infty$.
\end{itemize}

\begin{Remark}{\em (Pole of the path-loss function)}
The above  phase transition is not linked to the pole of the
simplified path loss function~(\ref{simpl.att}) used in the analysis.
To show  this, one can consider e.g. $l(u)=(\max(1,u))^4$ and
evaluate explicitly the integral in the left-hand-side
of~(\ref{e.PINR-interf})
%Using \ref{Ieq:chssarctan}) with $t=\mu(1-p)T l(r)$, we get that
%in the fixed Poisson receiver model,
%\begin{eqnarray*}
%\int_0^\inft \frac{pT l(r)}{l(v)+ (1-p)T l(r)} v\, \md v &  = & 
% \frac p {1-p}
%\left[-\frac{1}{2A^2}\sqrt{(1-p)T l(r)}
%\arctan\biggl((Ar_0)^2\sqrt{\frac{1}{(1-p)T l(r)}}\biggr)
%\right. \\ & & \left.+
%\frac{\pi}{4A^2}\sqrt{(1-p)T l(r)}+
%\frac{r_0^2}{2} \frac{(1-p)T l(r)+(Ar_0)^4(1-\mu^{-1}}{(1-p)T l(r)+(Ar_0)^4}\right]\,.
%\end{eqnarray*}
%Hence, since $l(r)= (Ar)^4$ for $r\ge r_0$, the dominant term in the
%last function is 
%$$ \exp\left[ \frac{\pi}{4}\frac{p}{\sqrt{1-p}} \sqrt{T} r^2 \right]\, ,$$
%when $r$ tends to $\infty$.
%Let us take the instance of the Poisson receiver model: we get 
% from~(\ref{e.delay-INR-main}) and~(\ref{eq:ch9basicld})
%%(\ref{eq:ch9el}) 
(we skip the details due to the lack of space) and conclude that 
that the mean delay is finite if $p\ne 0$ and
$$ \lambda_0> \lambda \hat \theta= \lambda \frac{\pi}{2}\frac{p}{\sqrt{1-p}} \sqrt{T} $$
and infinite if either $p=0$ or $ \lambda_0< \lambda \hat \theta$.
\end{Remark}
%In the dynamic split model for  OPL~1 with $\beta=4$,
%the mean delay is finite if $p\ne 0$ and
%$$ p(1-p)^{-3/2} < \hat \xi= \frac 4 { \sqrt{T}}.$$
%and infinite if $p= 0$ or $ p(1-p)^{-3/2} > \hat \xi$.

\begin{Remark}
Here are a few comments on this phase transition.%
\index{phase transition!wireless contention}
\begin{itemize}
\item The fact that $p=0$ ought to be avoided for having $\E^0[\LD] <\infty$
is clear;
\item The fact that intensity of potential transmitters 
$\lambda_0$ cannot be arbitrarily small when the other
parameters are fixed is clear too as this implies that:
(i)~at any given time slot,
the transmitters compete for too small set of receivers;
(ii)~targeted receivers are too far away from their transmitter. 
\item For $T$ and $\beta$ fixed, stability ($\E^0[\LD]<\infty$) requires that 
receivers outnumber potential transmitters by a factor which grows like 
$p(1-p)^{2/\beta-1}$ when $p$ varies; if this condition is not satisfied,
this drives the system to instability because
some receivers have too persistent interferers nearby (for instance, if $p=1$,
a receiver may be very close from a persistent transmitter which will
most often succeed, forbidding (or making less likely) the success of
any other transmitter which has the very same receiver). 
\end{itemize}
\end{Remark}
\paragraph{MNN model}
Fix $a,r\ge0$. For the path loss function~(\ref{simpl.att}) we have
\begin{eqnarray*}
\lefteqn{
\int_{ar}^\infty \frac{pT l(r)}{l(v)+ (1-p)T l(r)} v \,\md v}\\
&=&\frac 1 2 r^2 p (1-p)^{\frac 2 {\beta} -1} T^{\frac 2
  {\beta}}H(a,T(1-p),\beta/2)\, ,
\end{eqnarray*}
with
\begin{equation}
H(a,w,b) 
= \int_{a^2 w^{-1/b}}^\infty \frac 1 {1+u^{b}} \,\md u.
\end{equation}
Let
\begin{equation}
J(w,b) =  \int_{\theta=-\pi/2}^{\pi/2} 
H(2\cos(\theta),w,b)  \,\md \theta\, .
%\int\nolimits_{u> (2\cos(\theta))^2 w^{-1/b}} \frac 1 {1+u^{b}} \md u.
\end{equation}
From~(\ref{e.delay-MNN-main}) and~(\ref{eq:ch9basiclddsm}), 
we then get the same type of phase transitions as for the IPNR
model above:
\begin{itemize}
\item If $p\ne 0$ 
\begin{equation} 
\frac p{(1-p)^{1- \frac 2 \beta}}
T^{\frac 2 {\beta}} \left( \frac{K(\beta)} {2} + J(T(1-p),\frac \beta 2)\right)
< \pi,\end{equation}
then $\E^0[\LD] <\infty$;
\item If
\begin{equation}
\label{eqalo:dsminstab}
\frac p{(1-p)^{1- \frac 2 \beta}}
T^{\frac 2 {\beta}} \left(\frac{K(\beta)} {2} + J(T(1-p),\frac \beta 2)\right)
> \pi,
\end{equation}
then $\E^0[\LD] =\infty$.
\end{itemize}
We can use the bounds of Remark~\ref{remal:dsmbounds} to get the 
following and simpler conditions:
\begin{itemize}
\item If $p\ne 0$ and
$\theta (p,T,\beta) < \pi$
then $\E^0[\LD] < \infty$.
\item If  
$\theta (p,T,\beta) > 2\pi\,$
then $\E^0[\LD] = \infty$.
\end{itemize}

%\begin{Remark}
%\label{rem:alo-cone}
%It is easy to extend the results of
%Proposition~\ref{p:delay-NN-main} 
%to the case where one selects the closest receiver in a cone
%rather than the closest receiver in the whole space, for instance,
%considering in  the angle of the cone is $\pi$.
%then
%\index{nearest neighbor!in a cone}
%\begin{equation}\label{eq:ch9eldsm-cone}
%\E^0[\LD ]=\frac{2\pi\lambda_0}{p}\int_0^\infty re^{-\pi\lambda_0r^2}
%\calL_W(\mu Tl(r))\,\calD_I(r,Tl(r))\,
%\,\,\md r\,, 
%\end{equation}
%where  $\calD_W(s)$ is as in
%Proposition~\ref{p.delay-bipolar-fast-fading} and 
%\begin{eqnarray*}
%\calD_I(r,s)& = & 
%\exp \left\{ \pi \lambda \int_0^\infty \frac{ps}{l(v)+ (1-p)s} v\,\md v\right\}\\
%&&\times\exp \left\{2 \lambda \int_{\theta=0}^{\pi/4} \int_0^\infty \frac{r}{\cos(\theta)}
%\frac{ps}{l(v)+ (1-p)s} v\,\md v\,\md \theta\right\}\\
%&&\times \exp \left\{2 \lambda \int_{\theta=\pi/4}^{\pi/2}\,\,
%  \int_{r\cos(\theta)}^\infty \frac{ps}{l(v)+ (1-p)s} v\,\md
%  v\,\md \theta\right\}\,.
%\end{eqnarray*}
%We get the same phase transition phenomena as above, though with different 
%multiplicative constants. 
%This is easily extended to the MANET receiver model.
%The more acute the cone, the lower the thresholds (in the variable $p$) 
%which separate the two phases. 
%\end{Remark}

\begin{Remark}{\em (RESTART, cont.)}\label{R.restart-cont}
We continue the analogy with the RESTART algorithm described in Remark~\ref{R.restart}.
\index{RESTART algorithm}
Consider the fast Rayleigh fading, with slow, constant noise $W=Const$
in  the context of one of the nearest receiver models.
Then the local delay time of a packet can be seen as an instance of this
algorithm with the following identification:
$A_n=F^0_0(n)e_0(n)$ and $B=TW l({\cal D})$, where ${\cal D}$ is the (random) distance between
the node where the packet is located and the target receiver.
The support of $ l({\cal D})$ is unbounded (for instance, in the
Poisson receiver 
model, for all the considered path-loss  models, the density of ${\cal D}$ at $r>0$ is
$\exp(-\lambda_0\pi r^2) r$ for $r$ large).

The interference limited case can be seen as an extension of the
RESTART algorithm where the file size varies over time. More
precisely, the model corresponding to e.g. MNN is that where at attempt $n$,
the file size is $B_n=f(\Phi,C_n)$, where $\Phi$ is the Poisson p.p. and
$\{C_n\}_{n>0}$ is an independent i.i.d. sequence (here $C_n$
is the set of fading variables and MAC decisions at time $n$). 
\end{Remark}

\section{Finite Mean Delays and Diversity}
\label{s.FiniteMean}
As we saw in the last subsection, the existence
of big void regions as found in Poisson configurations leads 
to the surprising property in the nearest receiver models, that the
mean (time) local delay is finite everywhere 
but may have an infinite spatial average in rather classical scenarios.
We describe below a few ways of getting finite mean spatial average of
the mean local delay 
in fast fading scenarios. All the proposed methods rely on an
increase of {\em diversity}: more variability in fading, more
receivers, more mobility, more flexible (adaptive) coding schemes.
\index{diversity}
\subsection{Heavy Tailed Fading}
\paragraph{Weibull Fading}
Assume the path loss function~(\ref{simpl.att}) and deterministic $W>0$,  
Recall from the discussion after
condition~(\ref{eq:oppsufcondnoisenndir})
(cf. also~(\ref{eq:oppsufcondnoisennds}) that in this case 
the  mean local delay of the typical node 
is infinite in the IPNR and MNN model (due to the
noise constraint) 
if one has the (fast) Rayleigh fading.   
However if we assume that $F$ is Weibull of shape parameter $k$ i.e. 
$\Pro[F>x]=\exp(-(x/c)^k)$,
for some $c$, with $c$ and $k$ positive constants, then  
the condition $k<2/\beta$ is sufficient 
to have $\E^0[\LD] <\infty$ in the noise limited scenario. Indeed then 
$\Pro(F>l(r)WT)= \exp\left\{-(l(r)TW/c)^k\right\}\ge
\exp\left\{-(TW/c)^k (Ar)^{2-\epsilon}\right\}$,
for $r\ge 1/A$, and some  $\epsilon>0$. 
Therefore the finiteness of $\E^0[\LD]$ (with canceled
interference)
follows from the fact that 
the integral
$$
\int_{1/A}^\infty r \exp\left\{-\pi \lambda_0 r^2 +
(TW/c)^k (Ar)^{2-\epsilon} \right\} \,\md r
$$
is finite.
\paragraph{Lognormal Fading}
Assume now
$F$ is lognormal with parameters $(\mu,\sigma)$,%
\index{fading!lognormal} 
that is $\log (F)$ is ${\cal N}(\mu,\sigma^2)$ (Gaussian of mean $\mu$
and variance $\sigma^2$) and that
$W$ is constant. Using 
$$
\Pro(F>x)
% & = & \Pro\left(\frac{\log(F)-\mu}{\sigma}>\frac{\log(x)-\mu}{\sigma}\right)\\
%&=& \frac 1{\sqrt{2\pi}} \int_{\frac{\log(x)-\mu}{\sigma}}^\infty 
%\exp (-u^2/2) \md u\\
\sim 
\frac 1 {(\log(x)-\mu)/\sigma}
\exp\left(-(\log(x)-\mu)^2/2\sigma^2\right),
$$
when $x\to \infty$ and 
$$ \Pro(F>x) \ge \frac  {(\log(x)-\mu)/\sigma} {1 +(\log(x)-\mu)^2/\sigma^2}
\exp\left(-(\log(x)-\mu)^2/2\sigma^2\right)$$
for all $x>0$
%Since,
%$$ \sum_m \Pro(F\leq l(r)WT)^m = \frac 1 {\Pro(F>l(r)WT)}\, ,  $$
%we get that
%Thus we get
%\begin{eqnarray*}
%\E^0[\LD] & \le & B +
%\frac{2\pi \lambda_0}  p \int_{r, l(r)WT)> \mu+C } r \exp\left(-\pi \lambda_0 r^2 
%\right) \exp\left(\frac{(\log(l(r)WT)-\mu)^2}{2\sigma^2}\right)
%\frac{1 +\frac{\log(x)-\mu)^2}{\sigma^2}}{\frac{\log(x)-\mu)}{\sigma}}
%\md r\\ &< &\infty,
%\end{eqnarray*}
%where $B, C$ are finite positive constants.
one can show that $\E^0[\LD]<\infty$ in the noise-limited scenario
for IPNR and MNN model  
with the path loss function~(\ref{simpl.att})
(we skip the details). 

Let us conclude that a fading with heavier tails may be useful {\em
  in the noise limited scenario} in that the
mean delay of the typical node may be infinite for the Rayleigh case and finite
for this heavier tailed case.

\subsubsection{Networks with an Additional Periodic Infrastructure}
\index{network!with periodic infrastructure}
The second line of thoughts is based on the idea that extra receiver should be added
to fill in big void regions.
We assume again fast Rayleigh fading
in conjunction with the 
``Poisson + periodic'' independent receiver model.
In this receiver model we assume that the pattern of potential
receiver consists of Poisson p.p. and 
an additional periodic infrastructure.
Since  there is a receiver at distance at most, say,  $\kappa$ 
from every point,
the closest receiver from the origin is at a distance at most $\kappa$ and 
$$ \E^0[\LD] = \int_0^\kappa \calD_W(\mu Tl(r))\,
\calD_I^{INR}(Tl(r))\, D(\md r)\,,$$
where $\calD_W$, $\calD_I^{INR}$ are as in
Proposition~\ref{p:delay-NN-main}
and $D(\cdot)$ is the distribution function of the 
distance from the origin to the nearest receiver in
this model. 
This latter integral is obviously finite.

Notice that periodicity is not required here. The only important property
is that each location of the plane has a node at a distance which is
upper bounded by a constant.

\subsubsection{High Mobility Networks}
\label{secalo:HMN}
It was already mentioned in Example~\ref{exe.basic}
that if one can assume that the whole network is independently  re-sampled 
at each time slot (including node locations
$\Phi$, which is {\em not} our default option) --- an assumption which
can be justified when there is  a high mobility of nodes 
 ---
then  $\E^0[\LD ]=1/p_c<\infty$  provided $p_c>0$.
This observation can be refined in at least two ways:
\begin{itemize}
\item Assume the IPNR model, with fixed potential receives, 
and high mobility of MANETS nodes, i.e., with $\Phi=\Phi(n)$
i.i.d. re-sampled at each $n\ge1$. Assume also fast noise and
fading. 
Then  one can easily argue that 
\vbox{\begin{eqnarray*}
\lefteqn{\E^0[\LD ]}\\
&&\hspace{-3em} =
2\pi \lambda_0 \int_{r>0} r \exp(-\pi \lambda_0 r^2)
\frac{1}{p{\cal L}_W(\mu l(r)T) {\cal L}_{I^1}(\mu l(r)T)} \md r.
\end{eqnarray*}}
The finiteness of the last integral can be assessed using arguments
similar to those given above.
\item  Assume now the IPNR model, with i.i.d. potential
  receives $\Phi_0(n)$  and static MANET $\Phi$.
Assume also fast noise and fast fading. 
We found no closed form expression  for the mean local delay of the
typical node in this
case, however using some convexity arguments it can be shown that
%Then  one can easily argue that 
%\begin{eqnarray*}
%\E^0[\LD ] = \E^0\left[ 
%\frac 1{p 2\pi \lambda_0 \int_{0}^\infty r \exp(-\pi \lambda_0 r^2)  {\cal L}_W(\mu l(r)T) 
%\prod_{i\ne 0} \left(
%1-p +p \frac{1}{1+ T l(r)/l(|X_i|)} \right) \,\md r}\right].
%\end{eqnarray*}
%for the last expression.
%However, a convexity argument shows that the R.H.S. is bounded from above by
%$$ 2\pi \lambda_0 \int_{0}^\infty
%\frac {r \exp(-\pi \lambda_0 r^2)} {p {\cal L}_W(\mu l(r)T)}
% \E^0\left[ \frac 1{\prod_{i\ne 0} \left(
%1-p +p \frac{1}{1+ T l(r)/l(|X_i|)} \right) } \right]\,\md r\, .
%$$
%This latter expression is precisely equal to~(\ref{e.delay-INR-main}),
it is smaller that  the mean local delay of the typical node 
in the original IPNR model.%  considered in this paper.
\end{itemize}
\rem
An important remark is in order. In the  examples considered in this 
section, we perform (at least some part of) the  space average
{\em together} with the time average to get the mean local delay.
This operation, which makes sense in the case of high mobility (of
potential receivers, MANET nodes) more easily leads to a finite
mean local delay of the typical node.
In contrast, in the previous sections (case of static $\Phi$ and
$\Phi_0$)
  we perform the time average first 
and then the space average, and we get a different result, which can
for instance 
be infinite.

\subsubsection{Adaptive Coding and Shannon Local Delay}
%A more radical, than theses discussed above, 
%solution to make  the mean transmission delay finite
One may argue that if the mean delays are infinite in
the previously considered models, it is primarily because of the
{\em coverage logic},
where one transmits full packets at 
time slots when the receiver is covered at the
required SINR and where one wastes all the other time slots.
This results in a RESTART mechanism
(cf. Remark~\ref{R.restart} and~\ref{R.restart}),
which in turn explains why we have heavy tails and infinite means.
Adaptive coding
offers the possibility of breaking the coverage/RESTART logic:
it gives up with minimal requirements on SINR 
and it hence provides some non-null throughput at each time slot,
where this throughput depends on the current value of the SINR,
e.g. via Shannon's formula as briefly described in what follows.

%There is no difficulty extensing the scenario of Section~\ref{ssec:thru}
%to the time dimension exactly as in Section~\ref{ss.ST-scenarios}).
Let $\calT_0=\log(1+\text{SINR}_0)$ be the bit rate obtained
by the typical node $X_0$ under, Palm probability $\Pro^0$, at time
slot 0, where $\text{SINR}_0=\text{SINR}_0(0)$ is given by~(\ref{eq:SINR}).
It is natural to define the {\em Shannon local delay} of the typical 
node $X_0$ as 
$$\LD^{Sh}=\LD^{Sh}_0=
\frac{1}{p\E^0[\calT_0\,|\,\calS]}\,,$$
namely as the inverse of the time average of $\calT_0$ 
given all the static elements (cf. Section~\ref{s.TheLD}).
This definition is the direct analogue of that of the
local delay in the packet model.
Observe that  
\begin{eqnarray}\nonumber
\lefteqn{\E^0[\log (1+\text{SINR}_0)\,|\,\calS]}\\
&&=\int_0^\infty \Pro^0\{\,\log(1+\text{SINR}_0)>t\,|\,\calS\,\}
\,\md t \nonumber\\
&&=\int_0^\infty \Pro^0\{\,\text{SINR}_0>e^t-1\,|\,\calS\}\,\md t \nonumber\\
&&=\int_0^\infty \pi_c(e^t-1|\calS)\md t\, \label{e:Tviap}
\end{eqnarray}
where $\pi_c(v|\calS)$ is as  defined in (\ref{chalopic})
and where we made the dependence on $T=v$ explicit.
Consequently, we obtain
\begin{equation}\label{ss.TheLDSh} 
\E^0[\LD^{Sh}]=\E^0\biggl[\frac1{\int_0^\infty\pi_c(v\,|\,\calS)/(v+1)\,\md
  v}\biggr]\,.
\end{equation}

We now show two examples where $\E^0[\LD]=\infty$ but 
$\E^0[\LD^{Sh}]<\infty$. 

\paragraph{Bipolar Receiver model}
Consider the slow noise, fast Rayleigh fading scenario. 
We saw in Remark~\ref{R:wireless-contention-phase-trans-bipolar} that in this case,
for Poisson Bipolar, noise limited networks, 
a necessary condition for $\E^0[\LD]<\infty$ is that the noise $W$
has finite exponential moment $\E[e^{\{WTl(r)\mu}]<\infty$.  
For the mean Shannon local delay of the typical node 
we have 
$$\E^0[\LD^{Sh}]=
%\E\biggl[\frac1 
%{\int_0^\infty e^{-Wvl(r)\mu}/(v+1)\,\md v}\biggr]\\
\E\biggl[\frac{W} 
{\int_0^\infty e^{-vl(r)\mu}/(v/W+1)\,\md v}\biggr]\,,
$$
in the noise limited case.
It is easy to see that the last expression is finite provided
$\E[W]<\infty$ (which is much less constraining than the finiteness of
the exponential moment).

\paragraph{IPNR model}
Consider now the IPNR model, with fast Rayleigh fading.
Consider the interference limited case and with the
path loss function~(\ref{simpl.att})
It follows from the discussion after
Corollary~\ref{c:PINR-delay-interflim},
that if $\lambda_0\pi<\lambda\theta(p,T,\beta)$,
where $\theta(\cdot)$ is given by~(\ref{e:theta-delay}), then
$\E^0[\LD]=\infty$. 
For the Shannon delay, in the interference limited case, we have
{\small
\begin{eqnarray*}
\lefteqn{\E^0[\LD^{Sh}]=2\pi\lambda_0\frac{1}{p}\int_0^\infty
re^{-\lambda_0\pi r^2}}\\
&&\hspace{-2em}
%\E^0\biggl[\biggl(\int_0^\infty
%\frac{1}{v+1}\exp\Bigl\{\sum_{X_i\not=X_0}\log
%\calL_{eF}\bigl(\mu v(r/|X_i|)^{\beta}\bigr)
%%\Bigl(1-p+\frac
%%p{1+v(r/|X_i|)^\beta}\Bigr)
%\Bigr\}\,\md v
%\biggr)^{-1}\biggr]\,\md r\\
%&=&2\pi\lambda_0\frac{1}{p}\int_0^\infty
%re^{-\lambda_0\pi r^2}
\times\E^0\biggl[\biggl(\int_0^\infty
\frac{1}{v+r^\beta}\exp\Bigl\{\sum_{X_i\not=X_0}\log
\calL_{eF}\bigl(\mu v/|X_i|^\beta\bigr)
%\Bigl(1-p+\frac
%p{1+(v/|X_i|)^{\beta}}\Bigr)
\Bigr\}\,\md v
\biggr)^{-1}\biggr]\,\md r\,.
\end{eqnarray*}}
Using the inequalities
{\small \begin{eqnarray*}
\int_0^\infty\frac{\exp\{\dots\}}{v+r^\beta}\,\md v
&\ge&\frac{1}{2r^\beta}\int_0^{r^\beta}\exp\{\dots\}\,\md v
+ \int_{r^\beta}^\infty\frac{\exp\{\dots\}}{2v}\,\md v\\
&\ge&
 \min\Bigl(\frac{1}{2r^\beta},1\Bigr)
\int_0^\infty\min\Bigl(\frac1{2v},1\Bigr)\exp\{\dots\}\,\md v\\
&\ge&
 \min\Bigl(\frac{1}{2r^\beta},1\Bigr)
\int_0^{1/2}\exp\{\dots\}\,\md v\,.
\end{eqnarray*}}
Note that $\int_0^\infty 2r/(\min(r^{-\beta},2))e^{-\lambda_0\pi r^2}\,\md
r<\infty$. Using Jensen's inequality, we get that for all $X_i$
$$\log \calL_{eF}\bigl(\mu v/|X_i|^\beta\bigr)\ge
-\E[eF]\mu v/|X_i|^\beta=-p v/|X_i|^\beta$$ for $|X_i|>\rho$, 
where $\rho>0$ is some fixed constant. From this and from the inequality 
$\calL_{eF}\ge 1-p$ for $|X_i|\le\rho$,
we conclude that $\E^0[\LD^{Sh}]<\infty$ provided
$$
\E^0\biggl[\frac{\exp\Bigl\{-\log(1-p)\Phi(\{X_i:|X_i|\le\rho\})\Bigr\}}
{\int_0^{1/2}\exp\Bigl\{-pv\sum_{|X_i|>\rho}|X_i|^{-\beta}
\Bigr\}\,\md v}\biggr]<\infty\,.
$$
Using the independence property of the Poisson p.p., the fact that the
Poisson variable $\Phi(\{X_i:|X_i|\le\rho\})$ has finite
exponential moments, it remains to prove that
{\small 
$$
\E^0\biggl[\biggl(\int_0^{1/2}\!\!\!
e^{-(pv\sum_{|X_i|>\rho}|X_i|^{-\beta})}\,\md v\biggr)^{-1}\biggr]
=\E^0\Bigl[\frac{pJ}{1-e^{-pJ/2}}\Bigr]<\infty\,,
$$}
where $J=\sum_{|X_i|>\rho}|X_i|^{-\beta}$.
Note that for $J$ small, the expression under the expectation is close
to~2, whereas for $J$ bounded away from 0, we have 
\begin{eqnarray*}
\E^0\Bigl[\frac{pJ}{1-e^{-pJ/2}}\ind(J>\epsilon)\Bigr]
&\le& (1-e^{-p\epsilon/2})p\E^0[J]\\
&=&(1-e^{-p\epsilon/2})2p\pi\lambda\int_\rho^\infty t^{1-\beta}\,\md
t,
\end{eqnarray*}
which is finite since $\beta>2$. Note that the last inequality 
is essentially (modulo the problem of the pole of the 
simplified path loss function~(\ref{simpl.att}) at 0 equivalent to the
finiteness of the mean of the   
shot-noise. % is always assumed in our monograph (recall $\beta>2$).

\section{Conclusion}
\resetcounters
\label{s.conclusion}

In the present paper, we introduced a space-time scenario
for describing the dynamics of a MANET using Spatial Aloha.
This was used to analyze the law of the time to transmit a typical packet 
from a typical node to its next-hop node in such networks. 
This analysis was shown to lead to non trivial observations on the spatial
variability of the local delays in such MANETs, when assuming that the time
scale of the physical layer and the MAC layer is much smaller than
that of mobility. 
In this case, {\em the local delay of the typical node 
has heavy tails and infinite mean values}
in most standard scenarios, which however
{\em does {\em not} imply that the mean throughput of the typical node
is null}. 
In addition, a new kind of phase transition,
related to the  mean local delay of the typical node (being 
the spatial, large-population, average of mean delays experienced by
individual nodes) was identified for the
interference limited case; closed form expressions were also given for the 
thresholds separating the two phases in some computational cases.
Various ways of guaranteeing that the network is in the phase where
the spatial average of the mean delays is finite were discussed, some
based on an increase of 
the variability (very high mobility, heavy tailed fading), some based on 
bounding the distance to the next-hop. It was also shown that adaptive coding
offers fundamentally different performance compared to the coverage/outage
scheme, allowing for finite spatial mean local delays in cases 
when the non-adaptive coverage/RESTART scheme gives infinite values.    
The discussion on how to reduce the mean value of local delays (and in 
particular how to move from an infinite to a finite mean value)
opens many interesting research directions which are left for a companion
paper: for instance, it would be useful to understand what classes of
moderate mobility lead to such a decrease. The same question is natural 
for fading scenarios, or point processes representing additional receivers. 

\vspace{-2ex}
\singlespacing
\bibliographystyle{plainnat}
{\footnotesize 
\bibliography{delay}
}

\end{document}